\documentclass[10pt,conference]{IEEEtran}
\IEEEoverridecommandlockouts

\usepackage{cite}
\usepackage{amsmath,amssymb,amsfonts}
\usepackage{graphicx}
\usepackage{textcomp}
\usepackage{xcolor}
\usepackage{siunitx}
\usepackage{url}
\usepackage{tabularx}
\usepackage{subfig}
\usepackage{booktabs}
\usepackage{pgf}
\usepackage{pgfplots}
\usepackage{algorithm,algpseudocode}

\usepackage[mode=buildnew]{standalone}
\usepackage{tikz}
\usetikzlibrary{positioning}

\pgfplotsset{compat=1.18}

\begin{document}

\title{GenAI-Enhanced Digital Twins for Predictive\\
Interference Management in Ultra-Dense Networks}

\author{
\IEEEauthorblockN{Afan Ali, Ali Arshad Nasir and Daniel Benevides da Costa}
\IEEEauthorblockA{
Research Center for Communication Systems and Sensing (IRC-CSS)\\
Department of Electrical Engineering\\
King Fahd University of Petroleum and Minerals (KFUPM)\\
Dhahran, 31261, Saudi Arabia\\
\{afan.ali@kfupm.edu.sa,~anasir@kfupm.edu.sa,~danielbcosta@ieee.org\}
}
}

\maketitle

\begin{abstract}
Ultra-dense indoor next-generation networks suffer severe
interference from mobility-induced blockages and localized
multi-user hotspots that conventional digital twins~(DTs)
cannot anticipate. We propose a generative AI~(GenAI)-enhanced
DT framework employing a conditional generative adversarial
network~(cGAN) with a spatio-temporal generator and PatchGAN
discriminator for proactive rare-event channel synthesis. A
worst-case zero-forcing~(WC-ZF) beamformer driven by
Monte Carlo synthetic trajectories realizes distributionally
robust precoding, with control-channel overhead bounded to
$\approx$ 2.1\,kB per 10\,ms slot. Sionna-based simulations confirm substantial median SINR gains, large packet-loss reductions, and significant closure of the perfect channel state information (CSI) oracle gap, all achieved within a low inference overhead well inside the control-loop budget.
\end{abstract}


\section{Introduction}
\label{sec:intro}

Next-generation indoor wireless networks must deliver
terabit-per-second~(Tbps) throughputs, sub-millisecond
latency, and massive device connectivity for immersive
applications such as extended reality~(XR), holographic
telepresence, and industrial Internet of
Things~(IoT)~\cite{xue2024survey}. Operating in
millimeter-wave~(mmWave) and sub-terahertz~(THz) frequency bands
makes these targets achievable in principle, however, high
path loss, acute blockage sensitivity, and beam misalignment
compounded by dense small-cell deployments give rise to
highly dynamic interference patterns~\cite{kokkoniemi2025thz}.
Traditional interference management responds reactively to
instantaneous channel state information (CSI), consistently yielding significant packet-loss under changing
conditions~\cite{lohan2024survey}.

Digital twins (DTs) have emerged as a vehicle for proactive, AI-driven
network management by maintaining a continuously
synchronized virtual replica of the physical environment on
which AI/ML inference can be performed offline~\cite{deng2023dt,
becattini2024dt}. DTs, augmented with generative AI (GenAI), achieve
markedly better channel-modeling accuracy, accelerate deep
reinforcement learning~(DRL) convergence and reduce
decision-making latency~\cite{tao2024genai,
guan2025edge}. Despite these advances, existing DT frameworks
largely remain reactive, i.e., they cannot synthesize unseen
scenarios such as crowd-induced blockages or extreme
interference hotspots that are statistically rare but
operationally critical~\cite{chai2024genai,al2025pimrc}.
Embedding the generative model inside the DT, rather than
deploying it on the physical network directly, provides three
concrete advantages, i.e., the DT provides a site-specific, ray-traced propagation sandbox that 
conditions the GenAI on physically consistent channel distributions, 
accumulates rare-event radio frequency (RF) traces in the virtual domain without 
additional pilot overhead, and enables safe offline fine-tuning of 
both the DT calibration and the GenAI parameters through a closed 
feedback loop. These properties make the DT an architecturally 
essential component rather than an optional convenience, and they 
collectively motivate the framework proposed in this work.

\subsection{Contributions}

This paper proposes a GenAI-enhanced DT framework treating
the DT and the generative model as a single tightly coupled
system. Although GenAI-based DTs have begun to appear in the
literature~\cite{yang2025largemodel,basaran2025gentwin,
yang2025pimrc}, none
of the existing works simultaneously addresses proactive interference management with
an explicitly described generative
architecture, a closed-form worst-case
beamformer derived from synthetic
trajectories, and a quantified
synchronization overhead. The specific contributions are as follows:

\begin{itemize}

\item  Rather than treating the
channel model as a separate module, we
embed a geometry-based stochastic
channel model with explicit
material parameters for concrete,
glass, and wood at 73\,GHz, directly
into the DT via Sionna ray tracing.
This integration means that the GenAI module is
always conditioned on physically
consistent channel distributions.

\item We designed a GenAI module, which is built around a conditional generative adversarial network (cGAN), inside a DT feedback 
loop. To the best of our knowledge, no prior work employs closed-loop generative architecture for 
mmWave interference prediction. It synthesizes a distribution 
of physically plausible future interference maps, including 
rare blockage and hotspot events which are never observed by the 
physical network, at zero additional pilot cost.

\item The cGAN-predicted channel trajectories feed directly into 
a closed-form worst-case zero-forcing~(WC-ZF) beamformer 
derived analytically from $M$ Monte-Carlo samples, 
bypassing convex relaxation and semidefinite programming 
entirely, which is the 
first direct coupling of a generative channel predictor 
to a robust beamforming decision within a DT
control loop, transforming synthetic channel realizations 
into deployable beamforming vectors in a single forward 
pass.
\end{itemize}
The remainder of the paper is organized as follows: Section~\ref{sec:sysmodel} presents the system model. Section~\ref{sec:framework} details the proposed framework. Section~\ref{sec:results} evaluates performance, and Section~\ref{sec:conclusion} concludes with future directions.

\section{System Model}
\label{sec:sysmodel}

\begin{figure}[t]
  \centering
  \subfloat[]{%
    \includegraphics[width=0.46\columnwidth]{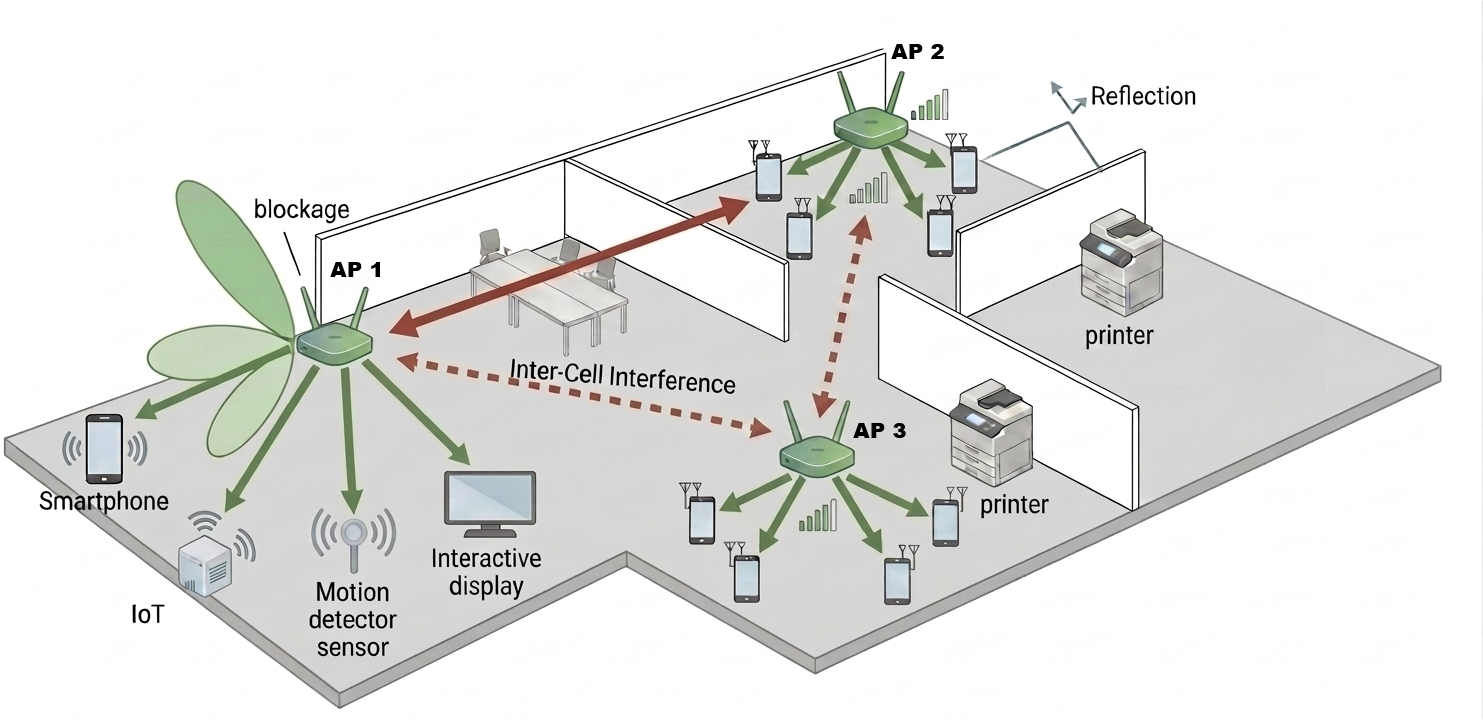}
    \label{fig:dense_campus}}
  \hfil
  \subfloat[]{%
    \includegraphics[width=0.46\columnwidth]{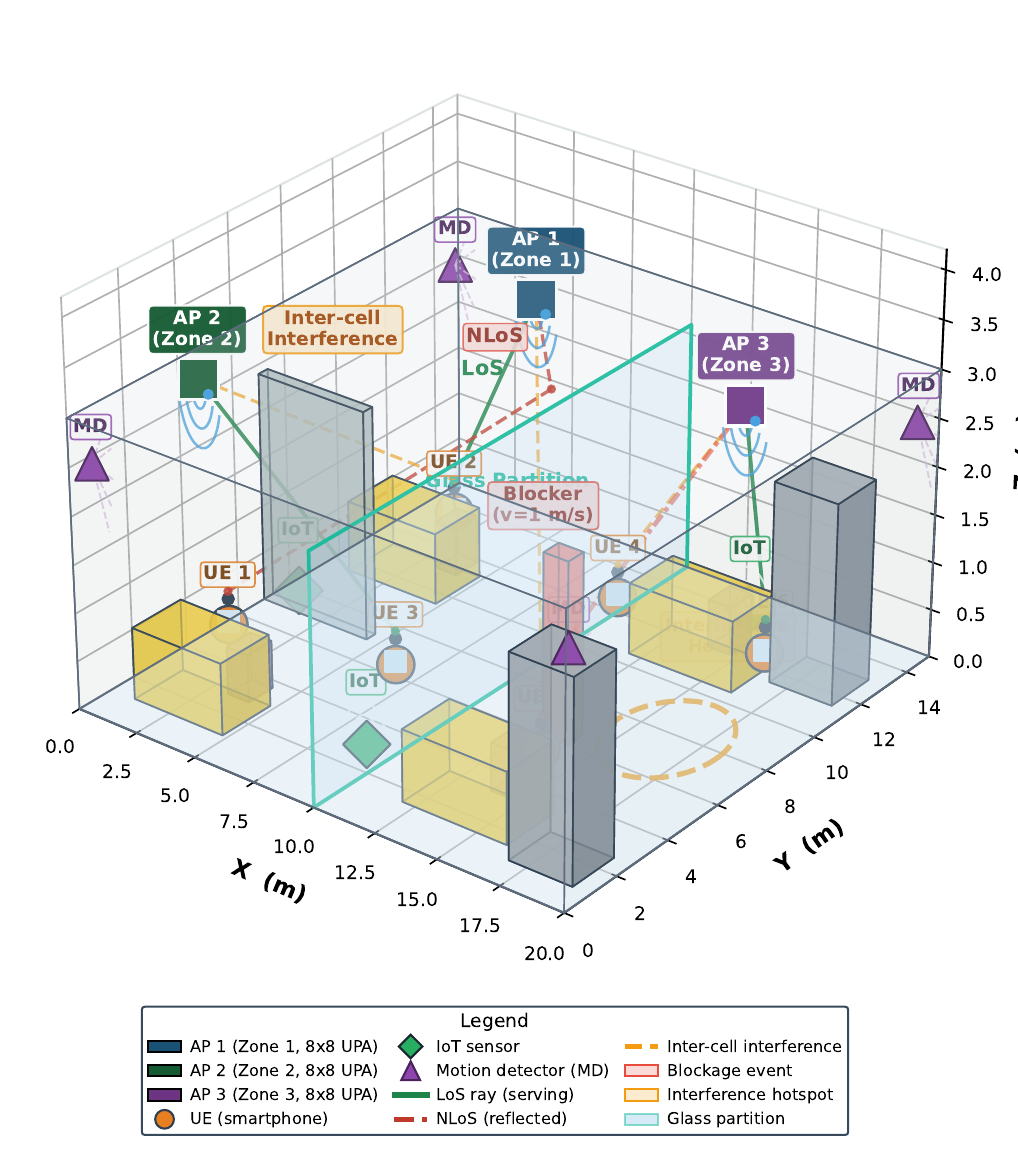}
    \label{fig:1b}}
  \caption{System model: (a)~dense indoor smart-campus
  deployment; (b)~simulated illustration for ray-tracing.}
  \vspace{-15pt}
  \label{fig:sysmodel}
\end{figure}

\subsection{Network and Channel Model}
\label{sec:network}

We consider a dense indoor next-generation network comprising
three small-cell access points~(APs), $A $, modeled after a smart-campus environment. 
As illustrated in Fig.\ref{fig:dense_campus}, a building inside smart-campus environment consists of multiple interconnected rooms and open workspace areas, where each AP is strategically deployed to provide coverage across spatially separated zones. To better visualize the propagation challenges within a smart campus, Fig.~\ref{fig:1b} presents a simulated view of the building, highlighting the presence of realistic physical obstacles, such as, concrete walls, wooden furniture, and glass partitions, along with the placement of three APs that must collectively manage severe multi-path scattering, blockage, and inter-zone interference. Collectively, the three APs serve $K $ heterogeneous user equipments~(UEs), including smartphones, IoT sensors, and motion detectors, across the dynamic open workspace~\cite{morabito2025genai}. Each AP carries $N_t$ transmit antennas arranged in an $8 \times 8$ uniform planar array~(UPA), and each UE is equipped with a single receive antenna. A central DT controller coordinates all APs via a 
dedicated control plane with a fixed synchronization 
period, $\Delta t_{\rm sync}\!=\!10$\,ms, enabling across all zones. Note that in this work, coordinated multi-point (CoMP) beamforming is specifically coordinated scheduling/beamforming, where each UE is served by a single AP but beamforming vectors are jointly optimized across APs to suppress inter-zone interference. UEs move under a 
random waypoint mobility model with 
$v_{\max}\!=\!1.0$\,m/s, reflecting realistic pedestrian dynamics within the campus environment.

The channel between AP~$b$ and UE~$k$ can be modeled using the
Sionna ray-tracing engine~\cite{sionna2022}, as a superposition
of $L_{k,b}$ multipath components, which can be defined as
\begin{equation}
  \mathbf{h}_{k,b} = \sqrt{\frac{N_t}{L_{k,b}}}
  \sum_{l=1}^{L_{k,b}}\alpha_{k,b,l}\,
  \mathbf{a}\!\left(\varphi_{k,b,l},\,
  \theta_{k,b,l}\right),
  \label{eq:channel}
\end{equation}
where $\mathbf{h}_{k,b} \in \mathbb{C}^{N_t \times 1}$,
the scaling $\sqrt{N_t/L_{k,b}}$ normalizes energy such that
$\mathbb{E}[\|\mathbf{h}_{k,b}\|^2] = N_t$ under the
assumption $\mathbb{E}[|\alpha_{k,b,l}|^2] = 1$, i.e.,
unit mean path gain per multipath component, $\alpha_{k,b,l} = \sqrt{PL_{k,b,l}^{-1}}\,e^{j\psi_{k,b,l}}$
is the complex path gain with ray-traced path loss
$PL_{k,b,l}$ and random phase
$\psi_{k,b,l} \sim \mathcal{U}[0,2\pi)$. The angles
$\varphi^T_{k,b,l}$ and $\theta^T_{k,b,l}$ are the
ray-traced azimuth and elevation angles of departure,
and the UPA steering vector
$\mathbf{a}_T(\varphi,\theta) \in \mathbb{C}^{N_t \times 1}$
has $(n_H, n_V)$-th element, which can be represented as
\begin{equation}
  \left[\mathbf{a}_T\right]_{n_H,n_V}
  = \frac{1}{\sqrt{N_t}}\,
  e^{\,j\frac{2\pi d}{\lambda}
  (n_H \sin\varphi\cos\theta + n_V\sin\theta)},
  \label{eq:steering}
\end{equation}
with $n_H \in \{0,\ldots,N_H\!-\!1\}$,
$n_V \in \{0,\ldots,N_V\!-\!1\}$,
$N_H = N_V = 8$ such that $N_t = N_H \times N_V = 64$,
half-wavelength spacing $d = \lambda/2$, 
and normalization $\|\mathbf{a}_T\|^2 = 1$.

\subsection{Signal Model}
\label{sec:signal}

Each UE~$k$ is served by AP $a(k) \in \{1,\ldots,A\}$,
selected via maximum received signal strength indicator (RSSI), partitioning all $K$ UEs into
disjoint zone sets $\{\mathcal{K}_b\}_{b=1}^{A}$, where
$\bigcup_b \mathcal{K}_b = \{1,\ldots,K\}$ and
$\mathcal{K}_b \cap \mathcal{K}_{b'} = \emptyset$,
$\forall\, b \neq b'$. The transmitted signal by AP~$b$ is given by
\begin{equation}
  \mathbf{x}_b = \sum_{j \in \mathcal{K}_b}
  \sqrt{P_{b,j}}\,\mathbf{w}_{b,j}\,s_{b,j},
  \label{eq:tx_signal}
\end{equation}
where $s_{b,j}$ is the unit-power data symbol,
$\mathbf{w}_{b,j} \in \mathbb{C}^{N_t \times 1}$ is the
beamforming vector with $\|\mathbf{w}_{b,j}\|^2 = 1$,
and $P_{b,j}$ is the allocated power. The received signal
at UE~$k$ can be written as
\begin{align}
  y_k &= \underbrace{
    \sqrt{P_{a(k),k}}\,\mathbf{h}_{k,a(k)}^H
    \mathbf{w}_{a(k),k}\,s_{a(k),k}
  }_{\text{desired signal}} \nonumber\\
  &+\underbrace{
    \sum_{\substack{b=1\\j\in\mathcal{K}_b\\
    (b,j)\neq(a(k),k)}}^{A}
    \sqrt{P_{b,j}}\,\mathbf{h}_{k,b}^H
    \mathbf{w}_{b,j}\,s_{b,j}
  }_{\text{interference}} + n_k,
  \label{eq:rx_signal}
\end{align}
where $n_k \sim \mathcal{CN}(0,\sigma^2)$ is additive white Gaussian noise (AWGN).
Compared to prior works that treat intra- and inter-cell
interference separately~\cite{morabito2025genai}, we unify
both into a single interference term, which allows a
compact worst-case signal-to-interference-plus-noise-ratio (SINR) formulation. Specifically, the
instantaneous SINR at UE~$k$ can be represented as

\begin{equation}
  \text{SINR}_k =
  \frac{P_{a(k),k}\,
  \bigl|\mathbf{h}_{k,a(k)}^H\mathbf{w}_{a(k),k}\bigr|^2}
  {\mathcal{I}_k + \sigma^2},
  \label{eq:sinr}
\end{equation}
where the unified interference term $\mathcal{I}_k$ can be defined as
\begin{equation}
  \mathcal{I}_k =
  \sum_{\substack{b=1\\j\in\mathcal{K}_b\\
  (b,j)\neq(a(k),k)}}^{A}
  P_{b,j}\,\bigl|\mathbf{h}_{k,b}^H
  \mathbf{w}_{b,j}\bigr|^2.
  \label{eq:interference}
\end{equation}
Under imperfect CSI, only $\hat{\mathbf{h}}_{k,b}$ is available rather
than the true channel $\mathbf{h}_{k,b}$. Moreover, the WC-ZF precoder
$\mathbf{w}_{a(k),k}$ is computed from $\hat{\mathbf{h}}_{k,a(k)}$, it
cannot coordinate against its own estimation error and the resulting
mismatch is, therefore, not inter-user interference but a perturbation
of UE $k$'s desired-link combining gain, which can be written as
\begin{equation}
I^{\text{err}}_k = \big|(\mathbf{h}_{k,a(k)} - \hat{\mathbf{h}}_{k,a(k)})^H
\mathbf{w}_{a(k),k}\big|^2.
\label{eq:Ierr}
\end{equation}
We conservatively bound this term by its full power and place it
alongside $\mathcal{I}_k$ in the SINR denominator, a worst-case choice
consistent with the minimax WC-ZF philosophy used throughout this
work, yielding a pessimistic SINR bound rather than treating
estimation error as interference. Substituting~\eqref{eq:Ierr}
into~\eqref{eq:sinr} gives the effective SINR under imperfect CSI as
\begin{equation}
\text{SINR}_k = \frac{P_{a(k),k}\big|\hat{\mathbf{h}}^H_{k,a(k)}
\mathbf{w}_{a(k),k}\big|^2}{\mathcal{I}_k + I^{\text{err}}_k + \sigma^2}.
\label{eq:sinr_imperfect}
\end{equation}

\subsection{Problem Formulation}
Three fundamental challenges arise in this setting:
\textbf{(C1)}~future CSI is unavailable at the
$\Delta t_{\rm sync}\!=\!10$\,ms decision instant;
\textbf{(C2)}~rare blockage events can collapse
${\text{SINR}}_k$ below a per-UE minimum SINR 
threshold $\gamma_k^{\min}$ within a single slot; and \textbf{(C3)}~imperfect CSI
inflates $I_k^{\rm err}$ during rapid transitions. To address all three simultaneously, we propose jointly 
optimizing the $\{\mathbf{w}_{b,k}\}$ and  
$\{P_{b,k}\}$ across all $K$ UEs and $A$ APs to 
maximize the sum spectral efficiency, represented as
\begin{align}
  \max_{\{\mathbf{w}_{b,k}\},\{P_{b,k}\}}\;&
  \sum_{k=1}^{K}\log_2\!\left(1 +
  {\text{SINR}}_k\right)
  \label{eq:problem}\\
  \text{s.t.}\quad
  &\|\mathbf{w}_{b,k}\|^2 \leq 1,\quad
  \forall\,k,\,b, \nonumber\\
  &{\text{SINR}}_k \geq \gamma_k^{\min},\quad
  \forall\,k, \nonumber\\
  &\sum_{j \in \mathcal{K}_b} P_{b,j} \leq P_{\max},\quad
  \forall\,b, \nonumber
\end{align}
over $M$ cGAN-generated synthetic channel
trajectories spanning a $T$-slot horizon. The
novelty lies in replacing the unknown true channel
$\mathbf{h}_{k,b}$ with a set of generative
realizations $\{\tilde{\mathbf{h}}_{k,b}^{(m)}\}_{m=1}^M$
drawn from the cGAN posterior, transforming
\eqref{eq:problem} into the distributionally 
robust program, represented as
\begin{equation}
\begin{split}
  \{\mathbf{w}_{b,k}^{\star}, P_{b,k}^{\star}\} =&\\
  \underset{\{\mathbf{w}_{b,k}\},\{P_{b,k}\}}{\arg\max}\;
  \underset{m \in \{1,\ldots,M\}}{\min}
  &\sum_{k=1}^{K}\log_2\!\left(1+
  {\text{SINR}}_k^{(m)}\right),
\end{split}
  \label{eq:wczf_obj}
\end{equation}
where $\mathbf{w}_{b,k}^{\star} \in \mathbb{C}^{N\times 1}$ 
is the optimal beamforming vector at BS $b$ for 
user $k$, $P_{b,k}^{\star}$ is the corresponding 
optimal transmit power, and 
${\text{SINR}}_k^{(m)}$ is the SINR of user $k$ evaluated on the $m$-th synthetic channel 
realization $\tilde{\mathbf{h}}_{k,b}^{(m)}$. To isolate the performance gain of the proposed GenAI-assisted worst-case beamforming strategy, $P_{b,k}^{\star}$ are assumed to be fixed throughout the optimization.
The inner minimization over $m$ enforces 
worst-case robustness across the cGAN posterior,
subject to the constraints in \eqref{eq:problem}. This worst-case selection directly resolves \textbf{(C1)} 
by acting on predicted channels before they are observed, 
\textbf{(C2)} by guaranteeing
${\text{SINR}}_k^{(m)} \geq \gamma_k^{\min}$
across all $M$ synthetic trajectories, and \textbf{(C3)} 
by replacing the imperfect estimate
$\hat{\mathbf{h}}_{k,a(k)}$ with cGAN-conditioned
samples $\{\tilde{\mathbf{h}}_{k,b}^{(m)}\}_{m=1}^{M}$
that encode the ray-traced multipath geometry,
thereby suppressing the residual leakage 
$I_k^{\rm err}$ in \eqref{eq:Ierr}. 

\begin{figure}[t]
 \vspace{2pt}
  \centering
  \includegraphics[width=1\columnwidth]{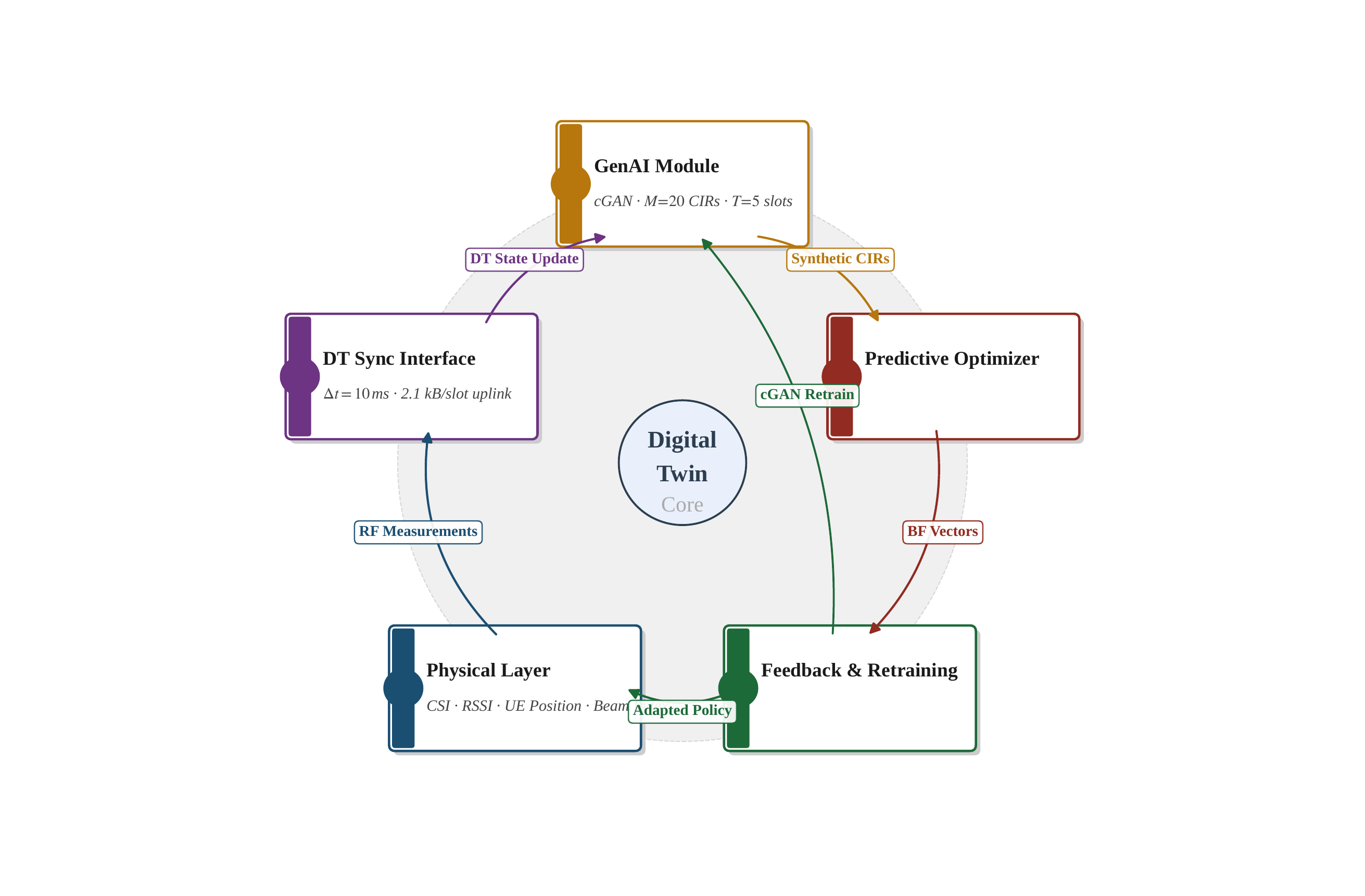}
  \caption{Closed-loop proposed architecture.}
  \vspace{-15pt}
  \label{fig:architecture}
\end{figure}

\section{Proposed GenAI-Enhanced Digital Twin Framework}
\label{sec:framework}

\subsection{Overall Architecture}
\label{sec:arch}

The proposed framework operates as a closed loop of five
interconnected modules, as illustrated in
Fig.~\ref{fig:architecture}. The physical layer
continuously collects RF measurements via pilot signals,
including CSI matrices, RSSI, beam indices, UE positions,
and traffic loads. The DT synchronization interface is triggered
every $\Delta t_{\rm sync}\!=\!10$\,ms, packaging three
data types per event: \textbf{(i)}~compressed CSI of
dimension $K\!\times\!N_t$ quantized at $B\!=\!4$\,bits
per complex component,
\textbf{(ii)}~RSSI and beam-index vectors of length $K$
with $b_{\rm RSSI}\!=\!8$\,bits per scalar, and
\textbf{(iii)}~UE position updates of size $3\!\times\!K$
with $b_{\rm pos}\!=\!16$\,bits per coordinate, where
$K\!=\!30$ UEs and $N_t\!=\!64$ transmit antennas.
The resulting uplink payload per slot can be computed as
\begin{align}
  \Omega &= \frac{K \cdot N_t \cdot B \cdot 2
  + K \cdot b_{\rm RSSI}
  + K \cdot 3 \cdot b_{\rm pos}}{8} \notag\\
  &= \frac{30\!\times\!64\!\times\!8
  + 30\!\times\!8
  + 30\!\times\!48}{8}
  \approx 2.1\,\text{kB/slot},
  \label{eq:overhead}
\end{align}
which, combined with ${\approx}$  $0.5$\,kB/slot downlink
beamforming updates, fits within the standard
control-channel budget without consuming the 10\,ms loop
period.

Upon DT state refresh, the GenAI module synthesizes $M$
plausible future channel impulse response (CIR) matrices over a $T$-slot
horizon, which are passed to the predictive interference
optimizer to derive proactive beamforming vectors. A feedback loop periodically
retrains the cGAN using measured SINR and packet-loss
data, where lightweight online updates every 100 slots and
full offline refinement every 1000 slots. At each
slot $t$, the DT executes a four-step cycle:
\textbf{(1)}~ingest synchronized physical-layer state;
\textbf{(2)}~generate $M$ synthetic CIR trajectories;
\textbf{(3)}~solve the WC-ZF beamforming problem; and
\textbf{(4)}~push per-AP beamforming matrices,
$\{\mathbf{W}^{\star}_b\}_{b=1}^{A}= 
[\mathbf{w}^{\star}_{b,1}, \ldots, 
\mathbf{w}^{\star}_{b,K_b}] 
\in \mathbb{C}^{N_t \times K_b}$, to all $A$ APs
via the downlink synchronization channel, achieving
CoMP precoding across all three deployment zones. Here $\mathbf{W}^{\star}_b$ simply stacks all per-UE vectors for AP $b$ into a single matrix for transmission via the downlink synchronization channel.

\subsection{GenAI Module}
\label{sec:genai}

The GenAI module is built around a cGAN, where a generator~$G$ 
and discriminator~$D$ are trained adversarially. Unlike a 
standard GAN, the cGAN conditions both $G$ and $D$ on a shared 
context vector, ensuring synthetic outputs remain physically 
consistent with the observed propagation environment.

\subsubsection{Generator Architecture}

The generator $G$ fuses two inputs: \textbf{(i)} a latent noise 
vector $\mathbf{z}\!\sim\!\mathcal{N}(\mathbf{0},\mathbf{I}_{128})$ 
for stochastic diversity, and \textbf{(ii)} a conditioning vector 
$\mathbf{c}\!\in\!\mathbb{R}^{d_c}$ aggregating three sources:

\textbf{(a) Channel embedding:} The last $\tau\!=\!10$ complex CIR 
matrices of size $K\!\times\!N_t$ are encoded by a three-layer 2-D 
convolutional neural network (CNN) with filters $\{64,128,256\}$ and kernel size~$3\!\times\!3$, yielding a 
256-dimensional embedding $\mathbf{e}\!\in\!\mathbb{R}^{256}$ that 
captures short-term channel variation 
patterns including Doppler shifts and incipient 
blockage signatures.

\textbf{(b) Per-UE RF scalars:} The RSSI and 
active beam index are appended for every UE, 
contributing $2K$ scalar RF 
measurements that capture instantaneous link 
quality and beam-alignment context, respectively.

\textbf{(c) Rare-event flag:} A binary scalar $f\!\in\!\{0,1\}$ 
indicates blockage or interference hotspot ($f\!=\!1$) versus 
normal operation ($f\!=\!0$), biasing $G$ toward worst-case 
scenarios during critical periods.
Concatenating these components gives the conditioning vector
$\mathbf{c}\!=\![\mathbf{e};\mathbf{r};f]\!\in\!\mathbb{R}^{d_c}$,
where $\mathbf{r}\!\in\!\mathbb{R}^{2K}$ is a 
vector stacking the RSSI and active beam index 
of each of the $K$ UEs, and $d_c$ is the total 
dimension of the conditioning vector $\mathbf{c}$, given by

\begin{equation}
  d_c = 256 + 2K + 1 = 317, \quad K = 30.
  \label{eq:dc}
\end{equation}

\subsubsection{Fusion and Up-sampling}
The fused vector $[\mathbf{z};\mathbf{c}]\!\in\!
\mathbb{R}^{d_z+d_c}$ is projected through a 
512-unit FC--ReLU layer, then up-sampled via three 
transposed-convolution blocks with filters 
$\{256,128,64\}$, stride~2, batch normalization, 
and ReLU activations. A final $\tanh$ layer maps 
outputs to $[-1,+1]$, rescaled to physical channel 
amplitudes, yielding a synthetic CIR tensor of 
shape $K\!\times\!N_t\!\times\!2$ (real and 
imaginary parts), with $N_t\!=\!64$.

\subsubsection{Discriminator Architecture}
$D$ follows the PatchGAN 
design~\cite{isola2017pix2pix}, adapted to CIR 
tensors by treating patches along the 
$K\!\times\!N_t$ ($30\!\times\!64$) dimensions as 
local antenna-user sub-arrays, enforcing spatial 
coherence across neighboring antenna elements and 
users. It comprises three strided-convolution 
layers with filters $\{64,128,256\}$, stride~2, 
and LeakyReLU (slope~$0.2$) activations to ensure 
stable gradient flow during adversarial training. 
A final $1\!\times\!1$ convolution produces a 
per-patch validity score map of shape 
$\lceil K/2^3\rceil\!\times\!\lceil N_t/2^3\rceil$, 
whose mean is passed to the WGAN-GP objective.

\subsubsection{Training with WGAN-GP}
Training solves the minimax  
objective~\cite{gulrajani2017wgan}, formulated as
\begin{align}
  \min_G \max_D \; \mathcal{L}(D,G) 
  &= \mathbb{E}_{x \sim p_{\rm data}}
  \bigl[D(x|\mathbf{c})\bigr] \nonumber\\
  &\quad - \mathbb{E}_{\mathbf{z} \sim p_z}
  \bigl[D(G(\mathbf{z}|\mathbf{c})|\mathbf{c})\bigr]
  \nonumber \\
  &\quad - \lambda_{\rm GP}\,\mathbb{E}_{\tilde{x}}
  \Bigl[\bigl(\|\nabla_{\tilde{x}} 
  D(\tilde{x}|\mathbf{c})\|_2 - 1\bigr)^2\Bigr],
  \label{eq:wgangp}
\end{align}
where $\lambda_{\rm GP}$ is the gradient 
penalty coefficient enforcing the Lipschitz 
constraint on $D$, and $\tilde{x} = \epsilon x + 
(1\!-\!\epsilon)G(\mathbf{z}|\mathbf{c})$, 
$\epsilon \sim \mathcal{U}[0,1]$, is a convex 
interpolation between real and generated samples. 
WGAN-GP approximates the Wasserstein-1 
distance~\cite{gulrajani2017wgan}, providing 
meaningful gradients even under disjoint support, beneficial for rare-event channel samples in 
our dataset. The connection to \eqref{eq:problem} 
is two-stage: $G$ first learns to generate 
$\tilde{\mathbf{h}}_{k,b}^{(m)}\!\approx\!
\mathbf{h}_{k,b}$, which are then substituted 
into \eqref{eq:wczf_obj} to solve the 
distributionally robust sum-rate maximization, 
where the quality of $G$ governs the tightness 
of the approximation. Both $G$ and $D$ are 
trained with Adam ($\text{learning rate}\!=\!2\!\times\!
10^{-4}$, $\beta_1\!=\!0.5$, 
$\beta_2\!=\!0.999$), with $D$ updated five 
times per generator step to ensure an accurate 
Wasserstein distance estimate~\cite{gulrajani2017wgan}.

\subsection{Worst-Case Proactive Beamforming}
\label{sec:opt}

\subsubsection{Interference Trajectory Prediction}

The cGAN generator produces $M$ synthetic CIR
trajectory sets over a $T$-slot horizon as
\begin{equation}
  \bigl\{\tilde{\mathbf{H}}^{(m)}_{t+1},\ldots,
  \tilde{\mathbf{H}}^{(m)}_{t+T}
  \bigr\}_{m=1}^{M},
  \label{eq:traj_set}
\end{equation}
where $\tilde{\mathbf{H}}^{(m)}_{\tau} \in
\mathbb{C}^{K \times N_t}$ is the stacked multi-user
channel matrix at future slot $\tau$ under trajectory
$m$, with rows $\tilde{\mathbf{h}}_{k,b}^{(m)} \in 
\mathbb{C}^{1 \times N_t}$.
The per-AP sub-matrix for AP~$b$ can be obtained by
row-selection, defined as
\begin{equation}
  \tilde{\mathbf{H}}^{(m)}_{b,\tau} =
  \boldsymbol{\Phi}_b\,
  \tilde{\mathbf{H}}^{(m)}_{\tau}
  \in \mathbb{C}^{|\mathcal{K}_b| \times N_t},
  \label{eq:submatrix}
\end{equation}
where $\boldsymbol{\Phi}_b \in
\{0,1\}^{|\mathcal{K}_b| \times K}$ is a binary
row-selection matrix that extracts the rows of
$\tilde{\mathbf{H}}^{(m)}_{\tau}$ corresponding
to UEs in $\mathcal{K}_b$, satisfying
$\boldsymbol{\Phi}_b\boldsymbol{\Phi}_b^T =
\mathbf{I}_{|\mathcal{K}_b|}$. These per-AP
sub-matrices are then passed to the WC-ZF
beamformer for minimax
optimization over all $M$ trajectories and
$T$ prediction slots.

\begin{figure*}[!t]
\vspace{0.2cm}  
  \centering
  \subfloat[CDF of post-beamforming SINR during rare events.]{%
    \includegraphics[width=0.26\textwidth]{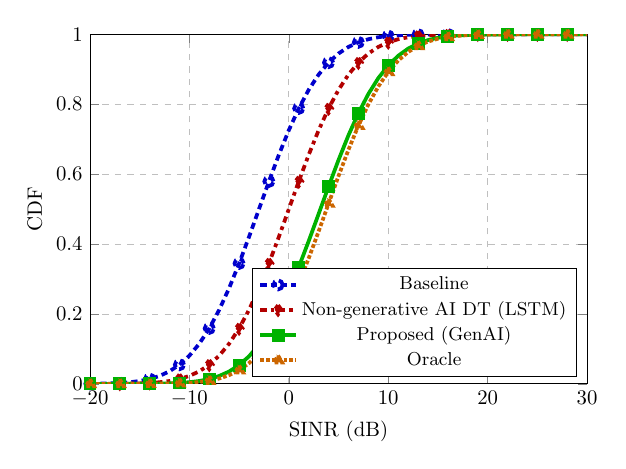}
    \label{fig:sinrcdf}}
  \hfill
  \subfloat[CDF of packet-loss rate per session.]{%
    \includegraphics[width=0.26\textwidth]{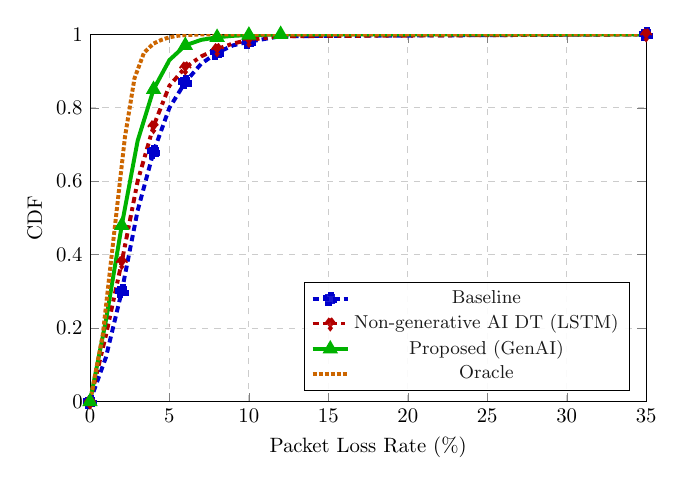}
    \label{fig:pktloss}}
  \hfill
  \subfloat[Post-beamforming SINR vs.\ number of APs.]{%
    \includegraphics[width=0.35\textwidth]{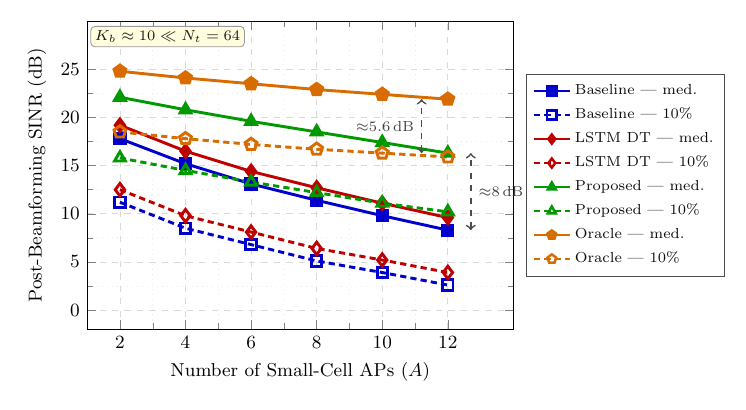}
    \label{fig:density}}
  \caption{Performance comparison of the proposed cGAN-DT
  framework against benchmarks.}
  \vspace{-15pt}
  \label{fig:results}
\end{figure*}

\subsubsection{Worst-Case Beamforming Selection}

The proactive beamforming vector $\mathbf{w}^{\star}_k$
for UE~$k$ served by AP~$a(k)$ is selected via a
minimax criterion over $M$ synthetic trajectories and
$T$ prediction slots can be illustrated as
\begin{align}
\mathbf{w}^{\star}_k &=
\arg\min_{\mathbf{w}}\;
\max_{\substack{m \in \{1,\ldots,M\}\\
\tau \in \{1,\ldots,T\}}}
\Bigl[-\,\text{SINR}_k\Bigl(\mathbf{w};\,
\tilde{\mathbf{H}}^{(m)}_{a(k),\tau},\,
\notag\\
&\phantom{=\arg\min_{\mathbf{w}}\;\max\;}
\bigl\{\tilde{\mathbf{H}}^{(m)}_{b,\tau}
\bigr\}_{b \neq a(k)}\Bigr)\Bigr],
\label{eq:wczf}
\end{align}
subject to $\|\mathbf{w}^{\star}_k\|^2 = 1$. The SINR evaluation in~\eqref{eq:wczf}
accounts for both intra- and inter-cell interference
from all $A$ APs as defined in~\eqref{eq:sinr}.
The minimax selection ensures that the achieved 
post-beamforming SINR at UE~$k$ satisfies 
$\gamma_k^{\min}$ under every synthetic 
trajectory and prediction slot, which can be written as
\begin{equation}
  \text{SINR}_k(\mathbf{w}^{\star}_k) \geq
  \gamma_k^{\min}, \quad \forall\, m,\,\tau,
  \label{eq:sinr_guarantee}
\end{equation}
This ensures robustness against the most adverse trajectory without convex relaxation or semidefinite programming. The final SINR is obtained by substituting $\mathbf{w}^{\star}_k$ into~\eqref{eq:sinr}, and when cGAN predictions are accurate, residual interference in~\eqref{eq:interference} remains near zero even after a blockage event.

\section{Simulation Setup and Results}
\label{sec:results}

\subsection{Simulation Setup}

Simulations use the Sionna ray-tracing 
library~\cite{sionna2022} in a 20$\times$15$\times$3\,m 
indoor office with concrete walls, with 0.2\,m thickness, glass 
partitions, wooden furniture, and metallic cabinets, with  
material parameters in Table~\ref{tab:params}. Three 
$8\!\times\!8$ UPA APs at 73\,GHz serve $K\!=\!30$ 
single-antenna UEs (${\approx}10$ per zone) under CoMP 
WC-ZF beamforming. The SBR tracer models LoS, 
reflections (up to 5 bounces), diffraction, and diffuse 
scattering. Two rare-event classes are injected: 
\textbf{(i)}~mobility blockages causing abrupt 
LoS-to-NLoS transitions ($p\!<\!0.05$ per slot, yielding on average 
one blockage event per 200\,ms of simulation); and 
\textbf{(ii)}~interference hotspots of 8--12 UEs within 
a 4\,m radius (10--15\% of traces). A 50\,000-step 
dataset is generated, each sample containing $\tau\!=\!10$ 
historical CIRs, future CIRs, SINR maps, and rare-event 
labels. A transmission is declared lost whenever \eqref{eq:sinr_guarantee}  is not satisfied and the packet-loss rate is the fraction of slots per session in which this occurs.

Four schemes are benchmarked: \textbf{(1)~Baseline}: 
reactive DT with no prediction, which applies ZF beamforming 
based on the most recently synchronized 
CSI, representing conventional 
interference management without any 
proactive capability; \textbf{(2)~LSTM DT}: 
two-layer LSTM (hidden size 256, Adam, 
learning rate$\!=\!2\!\times\!10^{-4}$, 800 epochs). The LSTM baseline's hyperparameters were selected via grid search chosen by lowest validation loss on a held-out 10\% split of the 50,000-step dataset, ensuring the comparison reflects a well-tuned rather than arbitrarily-configured baseline; 
\textbf{(3)~Proposed (GenAI)}: WC-ZF driven by $M\!=\!20$ 
cGAN trajectories; \textbf{(4)~Oracle}: perfect future 
CSI upper bound, which assumes exact knowledge 
of all $T\!=\!5$ future channel matrices 
and applies the same WC-ZF beamformer, 
thereby isolating the performance loss 
due to channel prediction error alone. Both the LSTM and cGAN are trained on the 
same 50\,000-step Sionna dataset to ensure 
a fair comparison. Scalability is assessed by varying 
deployed APs from 2 to 12 with $K\!=\!30$ UEs 
distributed uniformly. 

\begin{table}[t]
  \centering
  \caption{Simulation Parameters}
  \label{tab:params}
  \renewcommand{\arraystretch}{0.95}
  \setlength{\tabcolsep}{4pt}
  \begin{tabular}{@{}ll@{}}
    \toprule
    \textbf{Parameter} & \textbf{Value}\\
    \midrule
    Carrier freq.\ / Bandwidth  & 73\,GHz / 400\,MHz (3GPP NR FR2)\\
    AP array / UE antenna       & $8\times8$ UPA ($\lambda/2$) / Single\\
    Transmit power / Noise fig. & 23\,dBm / 9\,dB\\
    $\gamma_k^{\min}$ & 10 dB\\
    \midrule
    Concrete ($\varepsilon_r$, $\sigma$) & 5.31, 0.33\,S/m\\
    Glass ($\varepsilon_r$, $\sigma$)    & 6.27, 0.043\,S/m\\
    Wood ($\varepsilon_r$, $\sigma$)     & 1.99, 0.014\,S/m\\
    \midrule
    Ray-tracing / Max.\ reflections & SBR (Sionna) / 5\\
    APs $A$ / UEs $|\mathcal{K}_b|$ & 3 (one/zone) / ${\approx}10$\\
    Simulation steps / Mobility  & 50\,000 / RWP, $v_{\max}\!=\!1$\,m/s\\
    \midrule
    cGAN dim.\ $d_z$ / Batch/Epochs  & 128 / 64\,/\,500\\
    $\lambda_{\rm GP}$       & 10\\
    MC traj.\ $M$ / Horizon $T$      & 20 / 5 slots (50\,ms)\\
    \midrule
    Sync.\ $\Delta t_{\rm sync}$ / Payload & 10\,ms / 2.1\,kB/slot\\
    LSTM hidden size / Epochs          & 256 / 800\\
    \bottomrule
  \end{tabular}
\end{table}

\subsection{Results and Discussion}

\subsubsection{Post-Beamforming SINR under Rare Events}
Fig.~\ref{fig:sinrcdf} shows the cumulative distribution
function~(CDF) of post-beamforming SINR during rare
interference events. The proposed GenAI-DT achieves a median
SINR gain of 5--8\,dB over the baseline; the CDF rightward
shift confirms a substantial reduction in low-SINR outages.
The LSTM DT improves by 2--4\,dB over the baseline but
falls 5--8\,dB short of the proposed method, confirming
that distributional synthesis rather than point prediction
determines rare-event robustness. In the low-SINR tail, the
proposed method closes 60--70\% of the oracle gap.

\subsubsection{Packet-Loss Rate under Rare Events}
Fig.~\ref{fig:pktloss} shows the GenAI-DT achieves a
median packet-loss reduction of 60--70\% over the baseline.
The probability of packet-loss exceeding 5\% drops from 20\%
(baseline) to 15\% (LSTM DT) and further to 7\% (proposed),
closing 75--85\% of the oracle gap.

\subsubsection{Impact of Network Density}
Fig.~\ref{fig:density} shows post-beamforming SINR as the number of APs grows from 2 to 12, while maintaining the zero-forcing feasibility condition $K_b\!\approx\!10\!\ll\!N_t\!=\!64$ throughout. The conventional
baseline degrades most severely, where its median falls from 17.8 to 8.3\,dB and at
10\%-outage, it falls from 11.2 to 2.6\,dB. The LSTM DT provides modest
relief where its median falls from 19.2 to 9.6\,dB. The proposed GenAI-DT
remains most robust, i.e., median falls from 22.1 to 16.3\,dB, and with
10\%-outage, 15.8 to 10.2\,dB, reflecting the WC-ZF
beamformer ability to hedge against multiple simultaneous
interference sources.

\subsubsection{SINR Recovery After Blockage Events}

\begin{figure}[t]
  \centering
  \includegraphics[width=0.6\columnwidth]{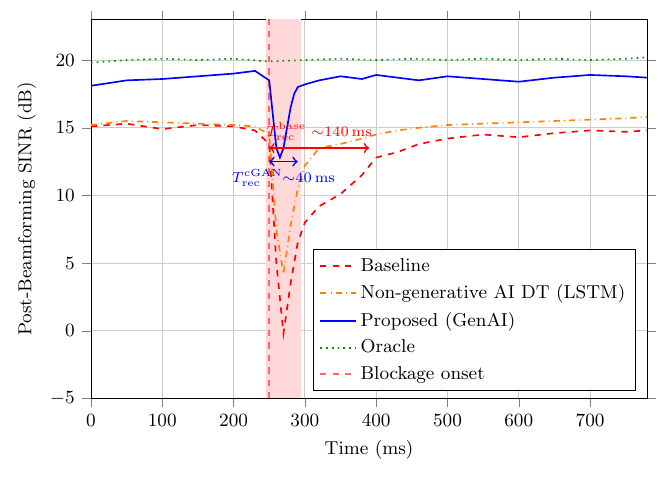}
  \caption{Post-beamforming SINR time trace around a sudden
  blockage event at $t=250$\,ms.}
  \vspace{-15pt}
  \label{fig:recovery}
\end{figure}

Fig.~\ref{fig:recovery} traces post-beamforming SINR around a blockage injected at $t=250$ ms. The baseline requires $T_{\rm rec}^{\rm base}\approx140$ ms to recover, as it must first detect degradation before recomputing beamforming vectors, while the LSTM DT reduces this to $\sim$90 ms but remains reactive. The proposed GenAI-DT commits a worst-case-robust beamforming vector proactively, limiting the instantaneous SINR drop to 8 dB and recovering to within 1 dB of the pre-blockage level in $T_{\rm rec}^{\rm cGAN}\approx40$ ms, a $3.5\times$ reduction in recovery latency over the baseline.

\subsection{Inference Latency and Complexity}
\label{sec:complexity}

All inference times are measured on a cloud GPU 
(PyTorch 2.0), representative of a cloud-based 
DT controller. The proposed GenAI-DT pipeline 
comprises a cGAN forward pass 
(${\approx}1.5$--$2.2$\,ms), $M$ 
Monte Carlo WC-ZF evaluations 
(${\approx}1.1$--$1.7$\,ms), and power 
normalization ($<\!0.2$\,ms), totaling 
$2.8$--$4.1$\,ms, within $41\%$ of the 
$10$\,ms control-loop budget. The remaining 
margin accommodates DT synchronization, 
uplink payload processing 
(${\approx}2.1$\,kB/slot), and feedback 
aggregation. Note that this refers to the 
control-loop latency, distinct from the 
sub-millisecond air-interface frame duration. 
Offline cGAN training is ${\approx}2.3\times$ 
more expensive than LSTM due to adversarial 
updates, but is performed once and 
fine-tuned periodically via the feedback loop.

\section{Conclusion}
\label{sec:conclusion}

This paper presented a GenAI-enhanced DT framework for
proactive interference management in dense indoor mmWave/sub-THz networks, where the DT and cGAN form a tightly coupled
system. Sionna-based simulations confirmed median SINR gains
of 5--8\,dB, packet-loss reductions of 60--70\%, and
60--85\% oracle gap closure within a 2.1--4.1\,ms inference
overhead, validating that generative synthesis with worst-case
robust beamforming effectively handles rare interference
events that reactive approaches miss. Future work will
explore interference-aware scheduling and federated multi-AP
coordination.

\section*{Acknowledgment}
The authors acknowledge the support provided by King Fahd University of Petroleum and Minerals (KFUPM), Dhahran 31261, Saudi Arabia, through the Interdisciplinary Research Center for Communication Systems and Sensing (IRC-CSS).

\bibliographystyle{IEEEtran}
\bibliography{Ref}

\end{document}